\documentclass[sigconf]{acmart}

\usepackage{colortbl}
\usepackage{array}
\newcolumntype{L}[1]{>{\raggedright\arraybackslash}p{#1}}

\AtBeginDocument{%
  }

\copyrightyear{2026}
\acmYear{2026}
\setcopyright{cc}
\setcctype{by}
\acmConference[CHI '26]{Proceedings of the 2026 CHI Conference on Human Factors in Computing Systems}{April 13--17, 2026}{Barcelona, Spain}
\acmBooktitle{Proceedings of the 2026 CHI Conference on Human Factors in Computing Systems (CHI '26), April 13--17, 2026, Barcelona, Spain}
\acmPrice{}
\acmDOI{10.1145/3772318.3790620}
\acmISBN{979-8-4007-2278-3/2026/04}

\begin{document}

\title{Envisioning Audio Augmented Reality in Everyday Life}

\author{Tram Thi Minh Tran}
\email{tram.tran@sydney.edu.au}
\orcid{0000-0002-4958-2465}
\affiliation{Design Lab, Sydney School of Architecture, Design and Planning,
  \institution{The University of Sydney}
  \city{Sydney}
  \state{NSW}
  \country{Australia}
}

\author{Soojeong Yoo}
\email{soojeong.yoo@sydney.edu.au}
\orcid{0000-0003-3681-6784}
\affiliation{Design Lab, Sydney School of Architecture, Design and Planning,
  \institution{The University of Sydney}
  \city{Sydney}
  \state{NSW}
  \country{Australia}
}

\author{Oliver Weidlich}
\email{oliver@contxtu.al}
\orcid{0009-0006-5907-2908}
\affiliation{
  \institution{Contxtual}
  \city{Sydney}
  \state{NSW}
  \country{Australia}
}

\author{Yidan Cao}
\email{yidan.cao@sydney.edu.au}
\orcid{0000-0003-3374-3524}
\affiliation{Design Lab, Sydney School of Architecture, Design and Planning,
  \institution{The University of Sydney}
  \city{Sydney}
  \state{NSW}
  \country{Australia}
}

\author{Xinyan Yu}
\email{xinyan.yu@sydney.edu.au}
\orcid{0000-0001-8299-3381}
\affiliation{Design Lab, Sydney School of Architecture, Design and Planning,
  \institution{The University of Sydney}
  \city{Sydney}
  \state{NSW}
  \country{Australia}
}

\author{Xin Cheng}
\email{xche7693@uni.sydney.edu.au}
\orcid{0009-0009-2812-528X}
\affiliation{Design Lab, Sydney School of Architecture, Design and Planning,
  \institution{The University of Sydney}
  \city{Sydney}
  \state{NSW}
  \country{Australia}
}

\author{Yin Ye}
\email{yiye0713@uni.sydney.edu.au}
\orcid{0009-0006-4098-5539}
\affiliation{Design Lab, Sydney School of Architecture, Design and Planning,
  \institution{The University of Sydney}
  \city{Sydney}
  \state{NSW}
  \country{Australia}
}

\author{Natalia Gulbransen-Diaz}
\email{natalia.gulbransen-diaz@sydney.edu.au}
\orcid{0000-0003-4810-1110}
\affiliation{Design Lab, Sydney School of Architecture, Design and Planning,
  \institution{The University of Sydney}
  \city{Sydney}
  \state{NSW}
  \country{Australia}
}

\author{Callum Parker}
\email{callum.parker@sydney.edu.au}
\orcid{0000-0002-2173-9213}
\affiliation{Design Lab, Sydney School of Architecture, Design and Planning,
  \institution{The University of Sydney}
  \city{Sydney}
  \state{NSW}
  \country{Australia}
}

\renewcommand{\shortauthors}{Tran et al.}

\begin{abstract}
While visual augmentation dominates the augmented reality landscape, devices like Meta Ray-Ban audio smart glasses signal growing industry movement toward audio augmented reality (AAR). Hearing is a primary channel for sensing context, anticipating change, and navigating social space, yet AAR’s everyday potential remains underexplored. We address this gap through a collaborative autoethnography (N=5, authoring) and an online survey (N=74). We identify ten roles for AAR, grouped into three categories: task- and utility-oriented, emotional and social, and perceptual collaborator. These roles are further layered with a rhythmic and embodied collaborator framing, mapping them onto micro-, meso-, and macro-rhythms of everyday life. Our analysis surfaces nuanced tensions, such as blocking distractions without erasing social presence, highlighting the need for context-aware design. This paper contributes a foundational and forward-looking framework for AAR in everyday life, providing design groundwork for systems attuned to daily routines, sensory engagement, and social expectations.

\end{abstract}

\ccsdesc[500]{Human-centered computing~Mixed / augmented reality}

\keywords{audio augmented reality, smart glasses, spatial audio, augmented sound, collaborative autoethnography}

\maketitle

\section{Introduction} 

Augmented reality (AR) has largely been driven by visual augmentation~\cite{dey2018systematic, zhou2008trends, kim2018revisiting, tran2023wearable}. While audio augmented reality (AAR) has historically been explored through headphones, earphones, and assistive technologies, recent advancements in audio-first smart glasses (like the Meta Ray-Ban\footnote{\url{https://www.meta.com/au/ai-glasses/}}), have expanded its form factor and functionality. These glasses prioritise audio as the primary interaction modality, but include cameras and contextual sensors to support greater environmental awareness and responsiveness; for example, Meta’s latest AI-integrated glasses can answer spoken queries and provide real-time translation. 

Despite growing industry interest, AAR remains significantly underexplored relative to visual AR~\cite{cagiltay2023case}. In most studies, sound is treated as a secondary output rather than a primary mode of engagement, or integrated only in niche applications such as assistive technologies (e.g., navigation for visually impaired users) and entertainment (e.g., gaming)~\cite{yang2022audio}. This leaves a significant gap in understanding AAR’s more generalised potential in daily life. Hearing is a primary channel through which people sense context, anticipate change, and navigate social space~\cite{mcadams1993thinking, oleksik2008sonic}. If we continue to overlook AAR in everyday settings, we risk designing spatial technologies that are visually rich but sensorially narrow.

This study takes \textbf{everyday life} as its central frame of reference. Drawing on Kossoff’s \textit{Domains of Everyday Life}~\cite{kossoff2015holism}, and an understanding of the everyday as the space where people meet needs and shape future possibilities, we examine how AAR might support existing routines. Lefebvre’s concept of rhythmanalysis~\cite{lefebvre2013rhythmanalysis} further highlights how daily life is structured by bodily, environmental, and social rhythms—many of them auditory—that AAR can enhance. We also build on Dourish’s view of embodied interaction~\cite{dourish2001action}, understanding sound not simply as output but as part of lived, socially situated activity.

By foregrounding the mundane and familiar, we aim to understand how AAR might enrich experience without competing for visual attention, particularly by offering forms of support that operate in parallel with ongoing tasks and environments. Attending to these ordinary moments also supports situated futuring, what \citet{sanchez2025future} describe as opportunities for profound learning about future technologies through engagement with day-to-day settings. This study is guided by the following research questions (RQs):

\begin{itemize}
    \item \textbf{RQ1}: How do people imagine AAR supporting or reshaping their everyday auditory experiences?
    \item \textbf{RQ2}: What boundaries or concerns do people identify when imagining AAR in everyday life?
\end{itemize}

To explore these questions, we adopt a dual-method approach conducted in parallel, with both methods addressing the same research questions through different modes of engagement. First, a collaborative autoethnography~\cite{chen2024towards, bala2023towards, chang2013collaborative} from five design researchers documented 72 reflections on their own everyday sound experiences, and asked them to imagine how these could be reshaped through audio augmentation. While autoethnography is inherently a first-person approach~\cite{kaltenhauser2024playing}, the collaborative form enables comparative reflection across multiple lived perspectives. Second, an online survey (N=74) gathered broader perspectives from a demographically diverse sample. The two methods complement each other: the autoethnography provides depth and situated insight into lived auditory experience, while the survey offers breadth and variation across everyday contexts. Together, they show how imagined uses, expectations, and concerns around AAR emerge across different forms of engagement.

\textit{Contribution Statement.} This study offers an empirically grounded and experience-led account of how people envision AAR in everyday life. We contribute a thematic analysis of the roles people imagine for AAR, revealing its function as a perceptual collaborator; and surface everyday tensions and concerns, including those related to control, unwanted automation, and privacy. Rather than presenting a technical prototype, we take a conceptual and futures-oriented approach, offering design groundwork for AAR systems attuned to everyday routines and social expectations. In doing so, we respond to recent calls in Human-Computer Interaction (HCI) to move beyond narrow or techno-centric futures~\cite{sanchez2025future}, contributing to a growing body of work that centres human experience and expands the design space of what AAR could become.



\section{Related Work}

\subsection{Foundations of Audio Augmented Reality} 

AR is inherently multimodal~\cite{schraffenberger2016multimodal, schraffenberger2018arguably}. While most AR research and applications have prioritised visual augmentation~\cite{dey2018systematic, zhou2008trends, kim2018revisiting, tran2023wearable}, recent work has started exploring other sensory modalities (including audio~\cite{kwok2019gaze}, haptics~\cite{ochiai2016cross}, olfactory~\cite{dmitrenko2016scent}, and even gustatory augmentation~\cite{narumi2011flavors}) to support richer and more immersive experiences~\cite{azuma1997survey, azuma2001recent, schraffenberger2016multimodal}. This paper focuses on AAR, an approach that overlays, filters, or enhances real-world sound through computer-generated audio to reshape everyday listening~\cite{cagiltay2023case, mariette2012human, yang2022audio}. 

AAR systems can be classified by how and where real-world (RW) and computer-generated (CG) sounds are mixed~\cite{lindeman2007classification}, including: \textit{\mbox{environmental} mixing}, where CG sound is played through speakers and blends with the RW space acoustics;  \textit{microphone-hear-through}, where RW sound is captured through microphones, processed, and digitally mixed with CG sound before being played through headphones; \textit{acoustic-hear-through}, which delivers CG audio through bone conduction while leaving the ears open to naturally receive RW sound.

The effectiveness of these different mixing methods depends not only on where sounds are combined~\cite{lindeman2007classification} but also on how much of the real-world audio is retained, a concept known as \textit{acoustic transparency}~\cite{mcgill2020acoustic}. Environmental mixing is fully transparent by default, as RW sound is not altered. In contrast, acoustic-hear-through systems (e.g., bone conduction) allow users to hear their surroundings naturally, while microphone-hear-through setups reintroduce RW sound after processing, which enables filtering but may distort timing or spatial cues.

While acoustic transparency enables a seamless blend of RW and CG sounds, noise isolation and cancellation do the opposite: blocking or suppressing RW audio to enhance immersion. Noise isolation relies on passive sound-blocking materials (e.g., earplugs, over-ear headphones), while active noise cancellation captures ambient sound and generates anti-noise signals to suppress it.

\subsection{Adaptive and Interactive Audio Augmented Reality}

To create seamless and immersive experiences, AAR systems must consider how CG sounds are positioned, how they adapt to the environment, and how users interact with them.

A foundational aspect of AAR is \textit{spatial audio}, which allows virtual sounds to be positioned in 3D space to match real-world acoustics and is widely seen as key to realism, immersion, and out-of-view awareness~\cite{blauert1997spatial, sundareswaran2003aar, kailas2021spatial}. Effective spatialisation typically relies on three components working together~\cite{yang2022audio}: (1) \textit{head-pose tracking}, which keeps spatial cues stable as the listener moves; (2) \textit{spatial-sound synthesis}, such as ambisonics~\cite{gorzel2019efficient} or binaural rendering, the latter commonly implemented with Head-Related Transfer Functions (HRTFs)~\cite{xie2013head} to support accurate localisation under head-direction–aware updates; and (3) \textit{room-acoustics modelling}, which uses techniques like artificial reverberation and real–virtual reverberation matching to ensure CG audio blends naturally with the surrounding environment~\cite{bona2022automatic}.

Equally important is \textit{context-awareness}, the ability of AAR systems to respond to surroundings, behaviours, or routines~\cite{meta2020audioresearch}. For example, systems may adjust audio levels based on ambient noise, prioritise certain sounds (e.g., speech), or adapt playback settings to location or activity~\cite{mynatt1997audio, bose_2018_audio}. Recent work has also explored how real-world sounds themselves can serve as inputs for AAR interaction. ~\citet{bhattacharyya2025birds}’s `sonic linking' system demonstrates how environmental audio events (e.g., bird calls, engine sounds) can trigger or modulate virtual behaviours, creating more responsive and tightly coupled audio-augmented experiences.

Interaction with AAR can be both \textit{explicit and adaptive}~\cite{yang2022audio}. Users can directly control playback, transparency levels, or sound selection through voice commands, touch gestures, or device inputs, ensuring flexibility across different environments~\cite{bose_2018_audio}. Beyond explicit control, head orientation-based interactions trigger contextual audio cues by detecting where the user is looking~\cite{bose_2018_audio, nassani2022attention}. Gaze-adaptive systems refine this further by analysing visual attention to dynamically adjust audio content~\cite{kwok2019gaze}. Eye movement tracking enhances this process, determining which surrounding sounds are most relevant and selectively amplifying or suppressing them based on user focus~\cite{meta2020audioresearch}.  

\subsection{Uncovering Meaningful Use Cases}

Recent commercial products, such as Meta Ray-Ban smart glasses and other open-ear wearables, have introduced AAR to mainstream consumers. These developments mark steps toward the broader vision of everyday or pervasive AR~\cite{bowman2021keynote, grubert2016towards}, where augmentation becomes embedded in daily routines and sensory engagement. In parallel, research has begun exploring how audio augmentation might support everyday activities. Identified use cases include multimodal overlays for learning musical instruments, real-time language translation via earpieces, and memory recall through recorded speech~\cite{mathis2024everyday}. Other research has examined preferences for altering or masking environmental noise~\cite{bustoni2024exploring} and the design of personalised audio spaces in homes that balance individual and social needs~\cite{jacobsen2024towards}. In contrast to these speculative and exploratory studies, a digital ethnography~\cite{tran2025wearable} investigating early adopters and technology reviewers of the Meta Ray-Ban documented actual usage patterns, with key use cases centring on media consumption (e.g., playing music), communication (e.g., phone calls), media recording, and interaction with the Meta AI assistant. While these studies provide valuable insights, none examine the comprehensive auditory experience of everyday life, considering both challenges to mitigate and opportunities to enrich personal and shared soundscapes. They have largely approached AAR through a problem-solving lens~\cite{mathis2024everyday, bustoni2024exploring} or within specific contexts (e.g., homes~\cite{jacobsen2024towards}), leaving open questions about its broader role in shaping daily auditory experiences. Therefore, our research aims to bridge this gap.

To identify meaningful use cases for AAR in everyday life, prior studies commonly adopt qualitative approaches, asking participants to reflect on daily experiences and how AAR could enhance them. For example, Mathis~\cite{mathis2024everyday} surveyed users about routine challenges before introducing AssistiveMR, a system featuring visual, auditory, and tactile augmentations, and asked them to envision its benefits. Similarly, \citet{jacobsen2024towards} organised design workshops to explore how technology could improve everyday listening, encouraging participants to reflect on home listening habits and shared auditory environments. Tran~\cite{tran2025everyday} used autoethnography to examine how existing technologies shape presence and engagement while speculating on the potential role of augmentation. A few studies take a more structured, quantitative approach. \citet{bustoni2024exploring} surveyed 124 participants on AAR strategies, investigating preferences for managing noise in different contexts, including whether users preferred to retain, modify, obscure, reduce, or remove sounds.

Building on these prior approaches, our study combines collaborative autoethnography (CAE) and an online survey to draw from both qualitative depth and broader quantitative perspectives. Autoethnography is a qualitative method based on first-person reflection, particularly well suited for gaining deep insight into the experiential and situated dimensions of technology use~\cite{kaltenhauser2024playing}. In HCI, its use has grown significantly, with 90.3\% (N=28) of published autoethnographies appearing in the last decade~\cite{kaltenhauser2024playing}, reflecting its increasing recognition and relevance. While this approach has gained traction in HCI venues, it remains largely absent from AR-related publications. CAE is a variant of autoethnography that emphasises dialogic interpretation across multiple researchers, offering both the depth of personal experience and the comparative richness of multi-perspective reflection~\cite{kaltenhauser2024playing, chen2024towards, bala2023towards}. As \citet{chang2013collaborative} argued, working with familiar self-data can lead to unchallenged assumptions, which collaboration helps to surface and refine. In HCI and design research, CAE is sometimes adapted to support experiential reflection and dialogic interpretation of technology use, rather than the full cultural critique typically associated with traditional autoethnography~\cite{kaltenhauser2024playing}. Our use of CAE follows this trajectory, focusing on collaborative reflection to surface differences and commonalities in everyday auditory experience.

For the online survey, we maintain the same objective of exploring AAR’s potential but extend the study by capturing diverse responses from potential users, ensuring that insights are not limited to expert reflection. This allows us to bridge deep, researcher-driven inquiry with broader, real-world perspectives, providing a more comprehensive understanding of how AAR might shape everyday auditory experiences.

\section{Methods}

\subsection{Collaborative Autoethnography}

\subsubsection{Researcher Backgrounds and Positionality} 

In this study, five researchers engaged in CAE, examining their personal encounters with everyday sound and speculating on how AAR might transform those experiences. This size was intentional: five participants allowed for a diversity of lived perspectives while remaining small enough to support meaningful dialogue and mutual interpretation. As such, this study represents a relatively large-scale autoethnographic effort, especially in HCI, where a recent systematic review found the average number of autoethnographers per study was just 1.7~\cite{kaltenhauser2024playing}.



The researchers shared an educational background in Interaction Design and HCI, but brought diverse research focuses (see \autoref{tab:positionality}). As design researchers, all were experienced in reflective and speculative design, uniquely positioned to critically analyse and interpret their own lived soundscapes. Their familiarity with AAR ranged from deep conceptual expertise to no prior exposure. This variation provided a valuable gradient of perspectives, helping to surface both informed design possibilities and curiosity-driven, first-time interpretations.

\begin{table*}[!ht]
    \caption{Overview of researchers’ research focus, everyday audio experience, and familiarity with AAR.}
    \centering
    \small
    \renewcommand{\arraystretch}{1.3} 
    \begin{tabular}{p{0.5cm}L{2.5cm}L{3cm}L{3cm}}
        \toprule
        \textbf{ID} & \textbf{Research Focus} & \textbf{Audio Experience} & \textbf{AAR Familiarity} \\
        \midrule
        R1 & Human-Vehicle Interaction & Music/concert lover; \newline frequent earphone use & High (conceptual expertise) \\
        R2 & Human-Robot Interaction & Music lover; plays casually; attends festivals & Moderate (related tech exposure) \\
        R3 & UX Design & Noise-sensitive; designed for voice UI & Moderate (related tech exposure) \\
        R4 & Multicultural Social Interaction & Amateur pianist; uses audio tools recreationally & None \\
        R5 & Trauma-Informed Technologies & Guided audio for \newline focus/relaxation & Moderate (conceptual awareness) \\
        \bottomrule
    \end{tabular}
    \label{tab:positionality}
\end{table*}

\subsubsection{Collaborative Autoethnography Process}

CAE was conducted over a three-week period (1–21 March 2025), following a preparatory week dedicated to setup and piloting. During this initial phase, the first author conducted a solo pilot autoethnography to test the data logging template and refine the reflection prompts. An onboarding session was held to familiarise all participating researchers with the study goals, key AAR concepts, and expectations for the documentation process. This session also served as an initial engagement point, helping researchers warm up and feel comfortable with the process of sharing personal experiences in later stages. In line with autoethnographic conventions, the `participants' in this dataset are the five researcher-authors, who contributed first-person reflections.

The CAE unfolded as a continuous and layered process, with individual reflections spanning three weeks, complemented by a mid-point discussion and a final collaborative synthesis session. These sessions were intentionally designed as collaborative reflection activities: researchers compared entries, examined differences in interpretation, and surfaced emerging patterns across individual accounts. In preparing these sessions, we considered elements commonly used in structured reflection design, such as role clarity, level of reflection (individual, parallel, collaborative), and the balance between synchronous and asynchronous engagement, to ensure that researchers could engage meaningfully and consistently. This approach aligns with core CAE principles~\cite{chang2013collaborative}, positioning self-observation within a shared analytic exchange rather than treating entries as isolated diary data.

\paragraph{\textbf{Individual Reflection (Weeks 1-3)}}

\noindent Researchers captured auditory experiences in their daily routines, noting both disruptions and moments of auditory meaning. The below prompts encouraged attention to sounds throughout the day, followed by deeper reflection in the evening. No quota of entries was required, researchers documented experiences whenever they felt notable. An example entry is shown in \autoref{tab:journal_format}, illustrating how experiences were recorded and interpreted in relation to AAR. For open science, the anonymised full dataset has been published on the Open Science Framework (OSF) (see \autoref{sec:studydata}).

\begin{table*}[ht]
    \caption{Example of a journal entry, with column headings reflecting the reflection prompts.}    
    \centering
    \small
    \begin{tabular}{L{2cm}L{5cm}L{5cm}}
        \toprule
        \textbf{Date \& Time} 
        & \textbf{What auditory experiences stood out to you today? Did you adjust or respond to sound in any way?} 
        & \textbf{Looking back, could AAR have helped, or would it have been unnecessary?} \\
        \midrule
        21 Feb 2025, 7:30 AM 
        &  
        I usually listen to music while waiting for coffee at a café, but I often have to take out my earphones or pause the music to make sure I hear when my name is called. I also tend to stay close to the counter so I don’t miss it.
        & 
        Maybe instead of lowering my music entirely while I wait, AAR could selectively amplify only the relevant sounds, like my name or my drink being called, while keeping everything else at normal volume. This way, I stay immersed in my own audio experience but don’t risk missing my order. \\
        \bottomrule
    \end{tabular}
    \label{tab:journal_format}
\end{table*}

\paragraph{\textbf{Midpoint Discussion (mid-Week 2)}}

\noindent Halfway through the reflection period, the team met to exchange entries and share early observations. While many initial reflections centred on moments of auditory discomfort that AAR might alleviate, the discussion also surfaced experiences where sound played a positive or meaningful role. This prompted a collective shift in emphasis for the final week, encouraging researchers to pay closer attention to such moments and consider how AAR might not only mitigate disruptions but also enhance valued aspects of auditory life. Some team members also expressed uncertainty about recording `ordinary' experiences that felt too mundane or universal. However, the group reaffirmed the value of interpreting familiar moments from diverse perspectives, recognising their potential to surface subtle but significant design insights.

\paragraph{\textbf{Collaborative Synthesis (end of Week 3)}}

\noindent By the end of the reflection period. The team reconvened for a closing discussion to reflect on the experiences captured. Each researcher selected one positive and one negative experience where AAR could be meaningfully leveraged, and shared these with the group. Others were invited to respond by drawing connections, highlighting contrasts, and speculating on alternative AAR possibilities. This discussion also produced a set of early codes, generated through a affinity-style grouping of similar reflections. These early codes then informed the first author’s thematic analysis.

\subsubsection{Data Analysis}
77 entries had been recorded. Five entries describing visual AR or unrelated scenarios were excluded, resulting in a final total of \textbf{72 entries} (Mean=14.4, Min=7, Max=19). 

An inductive thematic analysis~\cite{braun2006thematic} was conducted by the first author. The early codes from the collaborative session served as the starting point. Building on these, the first author conducted a full round of line-by-line coding across all entries to ensure that every instance of imagined AAR use was systematically captured. This process involved merging overlapping codes, separating codes that captured different ideas, and adding new codes as needed.

The analysis resulted in ten functional roles. Role labels were initially short two-word descriptors (e.g., `enhance/clarify,' `reduce/suppress') and were later streamlined into single-word titles for clarity. Definitions were sharpened to ensure that each role captured a specific and non-overlapping function. Each entry was assigned one or more roles based on how AAR was described as shaping the participant’s auditory experience.

\subsection{Online Survey}

\subsubsection{Survey Design and Distribution}

To complement the autoethnography, we conducted an online survey to identify broader patterns beyond the researcher cohort. The survey comprised four sections: everyday engagement with sound (Q2–Q6), speculative use cases for AAR (Q7–Q9), concerns or boundaries regarding its adoption (Q10–Q12), and demographic background (Q13–Q16). Q1 was a consent confirmation question.

Because AAR remains unfamiliar to many people, the survey included a short onboarding section before Q7 to introduce the concept. This provided a brief definition, examples of existing audio smart glasses (e.g., Meta Ray-Ban, Bose Frames), and a short explainer video outlining common features. The aim was to ensure that all participants shared a baseline understanding of AAR before responding to the speculative AAR questions.

The survey was hosted on Microsoft Forms and launched following an internal pilot with eight participants, which helped refine question clarity and estimate completion time (which was 10-15 minutes). This benchmark was later used to identify and exclude responses that appeared rushed (less than 5 minutes) or incomplete. The survey link was shared through the researchers’ social media networks (e.g., LinkedIn, Instagram) and physical posters across the University campus, primarily reaching individuals with limited prior experience with audio devices or smart glasses. To include perspectives from more technologically engaged users, the survey was also posted in sound- and AR-focused communities on Reddit and Facebook (see \autoref{sec:online}). The survey remained open for 28 days in March 2025.

\begin{table}[ht]
\caption{Survey participant demographics and AAR experience.}
\centering
\small
\renewcommand{\arraystretch}{1.3}
\begin{tabular}{@{}L{2.5cm}L{5.5cm}@{}}
\toprule
\textbf{Category} & \textbf{Details} \\
\midrule
\textbf{Age Range} & 21–70 years (M=37.8, SD=10.3) \\
\textbf{Age Groups} & 45\% (n=33): 25–34 \newline
30\% (n=22): 35–44 \\
\textbf{Device Ownership} & 89\%: In-ear earbuds \newline
81\%: Speakers (home/portable) \newline
59\%: Over-ear headphones \newline
15\%: Open-ear smart glasses \newline
8\%: Bone conduction headphones \\
\textbf{AAR Experience} & 41\%: No experience \newline
28\%: Read only \newline
12\%: Tried briefly \newline
8\%: Used occasionally \newline
9\%: Regular use \newline
1\%: Professional use \\
\textbf{Devices Mentioned} & Meta Ray-Ban (n=11), Bose Frames, Aftershokz, Halliday AR glasses (pre-ordered) \\
\textbf{Nationality} & American (15), Australian (13), British (13), German (8), Chinese (5), Vietnamese (2), Canadian, French, Nigerian, Bangladeshi, Belgian, British-Australian, Ghanaian, Indian, Iranian, Japanese, Norwegian, Palestinian, Romanian, Russian, Spanish, Turkish \\
\bottomrule
\end{tabular}
\label{tab:survey_demographics}
\end{table}

\subsubsection{Data Analysis}

A total of \textbf{74 valid responses} were collected. While the survey was not intended to be statistically representative, it complements the autoethnographic data by providing broader perspectives and helping to surface recurring themes and contrasts. \autoref{tab:survey_demographics} summarises participants’ demographics and prior experience with AAR.

To analyse the survey data, we used a mixed-methods approach combining descriptive statistics and qualitative thematic analysis. Quantitative responses, including Likert-scale ratings, multiple-choice selections, were summarised using frequency counts and visualised\footnote{\url{https://observablehq.com}} to identify broad trends. For open-ended responses, we applied inductive coding. Each response was extracted as a digital post-it note on an online Miro board\footnote{\url{https://miro.com/}}, then grouped into higher-level categories through iterative comparison and clustering. Questions 2–6 were not analysed, as they primarily served to support participants’ imaginative engagement with later questions. Our analysis focused on imagined AAR roles (Q7–Q9) and control preferences and concerns (Q10–Q12).

\section{Results}

This section reports findings from the CAE and survey separately. While structured by method, these results collectively address RQ1 (everyday visions) and RQ2 (concerns), which are synthesised in the Discussion.

\subsection{Collaborative Autoethnography}

\subsubsection{AAR Roles}

\begin{figure*}[ht]
  \centering
  \includegraphics[width=0.95\linewidth]{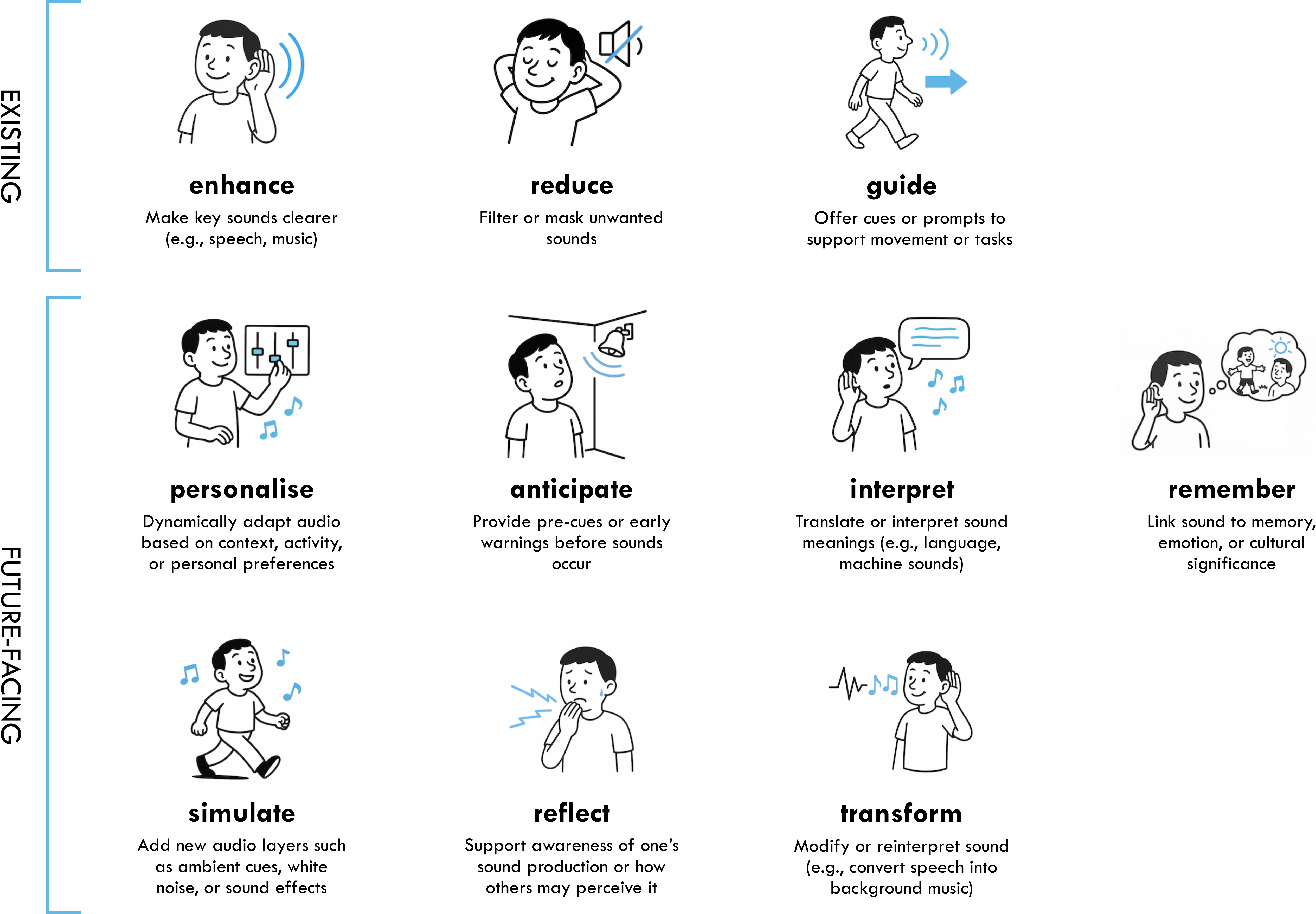}
  \caption{Thematic roles envisioned for AAR, organised into existing auditory practices and future-facing possibilities. Illustration credit: ~\autoref{sec:declaration}.}
  \label{fig:aar_roles}
  \Description{The figure illustrates nine thematic roles for Audio Augmented Reality (AAR) divided into “Existing” and “Future-facing” categories, each shown with a simple cartoon illustration, title, and brief description. Existing roles include enhance (making key sounds clearer, e.g., speech, music), reduce (filtering or masking unwanted sounds), and guide (offering cues or prompts to support movement or tasks). Future-facing roles include personalise (dynamically adapting audio based on context, activity, or personal preferences), anticipate (providing pre-cues or early warnings before sounds occur), interpret (translating or interpreting sound meanings, e.g., language, machine sounds), remember (linking sound to memory, emotion, or cultural significance), simulate (adding new audio layers such as ambient cues, white noise, or sound effects), reflect (supporting awareness of one’s sound production or how others may perceive it), and transform (modifying or reinterpreting sound, e.g., converting speech into background music).}
\end{figure*}

Our analysis revealed ten distinct roles that participants imagined AAR could play in everyday life (see \autoref{fig:aar_roles}). We grouped these roles into those reflecting existing auditory practices already supported by current technologies, and future-facing possibilities that extend beyond present capabilities. Existing roles were defined as those already supported by current audio technologies or common listening practices, such as speech-enhancement features (enhance), noise-cancelling and transparency modes (reduce), and task or navigation prompts (guide). This distinction underscores both the continuity with familiar listening practices and the imaginative potential of AAR to reshape everyday sound.

\textbf{Enhance} captures participants’ interest in how AAR might improve the clarity, intelligibility, and spatial awareness of everyday sounds. Across diverse scenarios, participants expressed a desire for systems that can selectively amplify relevant audio, or sharpen the perception of subtle but meaningful sounds. Some imagined AAR helping them hear conversations more clearly in public spaces, distinguish specific voices in group settings, or determine where a sound, like music or a voice, was coming from. As one participant put it, \textit{`When cooking, I wish AAR could have filtered out the cooking noise and made the TV audio clearer.'} These reflections highlight a recurring theme of perceptual support: using AAR to cut through auditory clutter, restore lost detail, and make soundscapes more comprehensible.

\textbf{Reduce} captures participants’ desire for AAR to minimise or mask intrusive, stressful, or contextually irrelevant sounds in daily life. From early-morning garbage trucks and snoring partners to loud strangers on public transport, participants frequently described situations where environmental sounds disrupted focus, rest, or emotional comfort. One participant recalled, \textit{`On the bus, a woman next to me kept talking loudly about political issues. Even loud music didn’t help me escape, if AAR could filter out this kind of unwanted talk or gently redirect my attention, that would be helpful.'} These reflections point to AAR’s potential to support psychological wellbeing and concentration by selectively reducing the salience of certain sounds.

\textbf{Guide} captures how participants envisioned AAR supporting everyday decision-making and safety through context-aware prompts and directional guidance. Some imagined AAR using spatial audio to reduce ambiguity in complex urban environments, such as intersections where closely positioned pedestrian signals produce overlapping beeps that can be difficult to interpret. Others suggested systems that could raise awareness of nearby vehicles or guide subtle adjustments in behaviour, such as prompting a user to step away from the road when traffic approaches from behind. Participants also envisioned personalised reminders tied to specific contexts: \textit{`AAR might detect that I am about to leave home and prompt: ``Do you want to confirm today’s home check?'''}, referring to appliances or doors that may have been left on or open. 

\textbf{Personalise} captures participants’ visions for AAR systems that adapt to individual preferences and situational contexts in real time. Across a range of everyday scenarios (e.g., commuting, working out, or relaxing), participants expressed a desire for audio experiences attuned to their mood, activity, location, and even physical condition. Some imagined adaptive systems that adjust volume or music genre based on environmental acoustics or biofeedback. One participant reflected that \textit{`when I’m not in a good physical condition, all of my preferences change,'} raising the question whether AAR could detect and respond to such subtle shifts. Another imagined the ability to \textit{`pick their own BGM [background music]'} in a group fitness class, allowing everyone to stay motivated while still keeping pace with shared movement and instructions. 

\textbf{Anticipate} highlights participants’ desire for AAR to provide advance warnings or modulation of sudden or important sounds. Several participants recalled moments when unexpected loud noises, such as a truck horn, fire alarm, or construction noise, startled or distressed them, prompting reflections on how AAR might offer pre-cues or gradual build-ups to reduce the shock. One participant recounted, \textit{`It was a [fire alarm] test that only lasted one or two seconds, but it was so loud and piercing'}. Others imagined AAR supporting alertness in everyday routines, such as reminding users of incoming deliveries: \textit{`I was waiting for the notification of food delivery. I find it a bit disturbing because I have to stay close to my phone.'} Across these instances, participants articulated a tension between needing to stay attentive to important auditory cues and the emotional or cognitive burden of being constantly on edge, suggesting that AAR could help mediate this balance through anticipatory design.

\textbf{Interpret} captures participants’ reflections on how AAR could make sense of ambient or environmental sounds that are difficult to identify or understand. Several participants described moments where they encountered ambiguous or unfamiliar sounds, such as a subtle mechanical issue in a dishwasher, a clicking noise in a moving car, or a low rhythmic beat from a neighbour, and expressed a desire for AAR to help interpret these sounds in real time. One suggestion imagined AAR-enabled vehicle systems that could detect and interpret abnormal sounds, providing specific feedback such as \textit{`Possible loose wheel component detected.'} In these cases, participants saw potential for AAR to reduce uncertainty and anxiety by offering contextualised interpretations or forwarding sound data for further diagnosis. Across these examples, participants valued AAR’s potential not only to enhance what is heard, but to clarify what it means.

\textbf{Remember} captures how participants reflected on the emotional and mnemonic qualities of sound, and how AAR might deepen or rekindle these connections. Several entries described everyday moments, such as hearing cicadas, arcade music, or the MS Teams ringtone, that unexpectedly evoked strong memories of childhood, cultural roots, or past digital experiences. One participant shared that the MS Teams calling tone \textit{`reminded me of when modems used to connect to the internet back in the early 2000s.'} Others imagined how AAR could intentionally build on this potential, for instance by overlaying ambient soundscapes to recreate a sense of place: \textit{`Background music and Cantonese voices transported me back to memories of Hong Kong.'}

\textbf{Simulate} sounds captures how participants imagined AAR adding new sonic layers to enhance or transform everyday experiences. Some participants envisioned AAR simulating natural soundscapes, such as context-aware white noise that adapts to the environment to sound as it is \textit{`supposed'} to. Others explored more playful applications, such as gamifying mundane tasks by adding sound effects to everyday actions (e.g., a `ding' sound when putting down a cup) to make daily life feel more engaging. One participant reflected on educational possibilities, proposing that AAR could be used to teach children about animal calls by simulating natural environments in urban classrooms or homes. 

\textbf{Reflect} captures participants’ awareness of how their use of sound technologies intersects with social presence and perception. Several participants reflected on moments of self-consciousness or uncertainty, such as wondering how loud their chewing sounded during a quiet film, or whether cleaning while wearing headphones might unintentionally disturb their neighbours. One recalled, \textit{`During quiet moments in the cinema, I became hyper-aware of chewing, it sounded extremely loud to me due to bone conduction.'} These moments of doubt often stemmed from a lack of awareness of how one’s own sounds are perceived by others, especially when audio devices mask external feedback. Some participants imagined AAR systems that could support perspective-shifting; for example, allowing users to hear how they sound to others, or helping them gauge how much noise they’re making while listening to music. Others considered how the form factor of audio devices affects social interpretation, noting that visible indicators like headphones may influence whether one appears approachable or attentive. 

\textbf{Transform} captures participants’ interest in AAR’s potential to replace unpleasant, stressful, or irritating audio with more soothing or contextually appropriate alternatives. Unlike suppression, which aims to remove or reduce unwanted noise, this role reflects a desire for auditory substitution: turning dentist drill sounds into calming audio, or replacing early morning garbage noise with birdsong that gradually increases in intensity as a natural wake-up cue. One participant suggested layering over disruptive bass from a neighbour’s music with gentle white noise, noting, \textit{`It’s not about filtering the sound out, it’s just that this exact low bass frequency or whatever it’s called really annoys me.'}

\subsubsection{Integration Concerns}
\label{tensions}
While participants proposed imaginative and often playful applications of AAR, they also surfaced important concerns about its integration into everyday life. Participants did not express concerns as simple `risks' but as tensions or negotiations between competing experiential values.

\begin{itemize}
    \item \textbf{Presence vs. peace} \\
    \emph{`Completely eliminating [a voice] might feel unnatural.'} \\
    This captures the need for quiet vs. the acknowledgment of another person’s presence. Participants don’t just want to mute others, they consider the emotional impact of doing so.

    \item \textbf{Safety vs. softness} \\
    \emph{`[...] reduces the shock factor while still keeping the car horn remains alarming.'} \\
    Rather than assuming all loud sounds are bad, participants distinguish between the function of a sound (safety) and its emotional impact (startle, fear). 

    \item \textbf{Intimacy vs. intrusion} \\
    \emph{`spatially adjusting the perceived position of voices to make storytelling even more intimate (could also feel quite creepy).'} \\
    Participants are drawn to personalised or spatialised sound but articulate where it crosses the line, showing sensitivity to how technology shapes emotional closeness and distance.
    \item \textbf{Artificiality vs. authenticity} \\
    \emph{`It could be great if I could adjust my own mix during live concerts… though it might feel a bit artificial.'} \\
    Participants welcomed the idea of customising audio environments, like isolating vocals or rebalancing instruments, but also questioned whether such control might compromise the rawness and spontaneity of live experiences.

    \item \textbf{Privacy vs. helpfulness} \\
    \emph{`While augmented sound should not be used to invade privacy or monitor private conversations [...]'} \\
   Participants envisioned AAR offering real-time translations or summaries to support inclusivity in public spaces, but these features must not come at the cost of privacy or surveillance.
\end{itemize}

\subsection{Online Survey}

\subsubsection{AAR Roles}

To enable comparison across the two datasets, we applied the same set of roles derived from the CAE to the open-ended survey responses. The survey yielded 139 AAR-related role count, compared to 76 from the CAE. This higher count reflects the inclusion of two open-ended questions in the survey, one focused on practical value (Q8), and the other on exciting or enjoyable possibilities (Q9). As there was substantial thematic overlap, we analysed responses to both questions together. To account for the difference in the data volume, we report both raw counts and proportional distributions (see \autoref{tab:aar_percentages}). The \textit{Combined} columns reflects a simple proportion of total coded instances across both datasets (N=215), rather than a weighted average. 

\begin{table*}[ht]
\caption{AAR role counts and proportional distributions across the CAE, online survey, and their combined totals. Colour shading in the `Total' column highlights the top (red), mid-range (orange), and lowest (yellow) roles.}
\centering
\small
\begin{tabular}{p{1.8cm}p{1.6cm} p{0.1cm} p{1.8cm}p{1.6cm} p{0.1cm} p{1.8cm}p{1.6cm}}
\toprule
\multicolumn{2}{c}{\textbf{CAE (N=76)}} && \multicolumn{2}{c}{\textbf{Survey (N=139)}} && \multicolumn{2}{c}{\textbf{Combined (N=215)}}\\
\cmidrule(r){1-2} \cmidrule(lr){4-5} \cmidrule(l){7-8}
\textbf{Role} & \textbf{Count–\%} && \textbf{Role} & \textbf{Count–\%} && \textbf{Role} & \textbf{Count–\%} \\
\midrule
\textbf{enhance}      & 13 (17.1\%) && \textbf{reduce}       & 38 (27.3\%) && \cellcolor{red!30}\textbf{reduce}       & \cellcolor{red!30}47 (21.9\%) \\
anticipate            & 11 (14.5\%) && personalise           & 32 (23.0\%) && \cellcolor{red!30}personalise          & \cellcolor{red!30}40 (18.6\%) \\
\textbf{reduce}       & 9 (11.8\%)  && \textbf{enhance}      & 24 (17.3\%) && \cellcolor{red!30}\textbf{enhance}      & \cellcolor{red!30}37 (17.2\%) \\
simulate              & 9 (11.8\%)  && interpret             & 13 (9.4\%)  && \cellcolor{orange!30}interpret          & \cellcolor{orange!30}21 (9.8\%) \\
personalise           & 8 (10.5\%)  && \textbf{guide}        & 12 (8.6\%)  && \cellcolor{orange!30}\textbf{guide}       & \cellcolor{orange!30}18 (8.4\%) \\
interpret             & 8 (10.5\%)  && anticipate            & 8 (5.8\%)   && \cellcolor{orange!30}anticipate          & \cellcolor{orange!30}19 (8.8\%) \\
\textbf{guide}        & 6 (7.9\%)   && simulate              & 6 (4.3\%)   && \cellcolor{orange!30}simulate           & \cellcolor{orange!30}15 (7.0\%) \\
remember              & 5 (6.6\%)   && transform             & 4 (2.9\%)   && \cellcolor{yellow!30}transform          & \cellcolor{yellow!30}7 (3.3\%) \\
reflect               & 4 (5.3\%)   && remember              & 1 (0.7\%)   && \cellcolor{yellow!30}remember           & \cellcolor{yellow!30}6 (2.8\%) \\
transform             & 3 (3.9\%)   && reflect               & 1 (0.7\%)   && \cellcolor{yellow!30}reflect            & \cellcolor{yellow!30}5 (2.3\%) \\
\bottomrule
  \addlinespace
  \multicolumn{8}{p{13cm}}{\textbf{Bolded roles} (enhance, reduce, guide) denote roles that correspond to well-established auditory practices.}

\end{tabular}
\label{tab:aar_percentages}
\end{table*}

\textit{Highest.} The most frequently identified role across both methods was \textbf{reduce} (21.9\% total mentions), reflecting a strong desire to remove or mask intrusive, irrelevant, or emotionally taxing sounds in daily life. This role appeared much more often in survey responses (27.3\%) than in autoethnographic entries (11.8\%), suggesting that broader users are particularly drawn to AAR’s potential to support comfort and relief.

\textit{Second Highest.} \textbf{personalise} ranked second overall (18.6\%), with participants expressing a clear interest in tailoring their auditory environment to suit mood, setting, and activity. This role was also more dominant in survey responses (23.0\%) than in the autoethnography (10.5\%), underscoring the appeal of having control and flexibility in everyday listening contexts.

\textit{Third Highest.} \textbf{enhance} (17.2\%) was well represented across both methods (17.3\% survey, 17.1\% CAE). Participants described wanting AAR to amplify important sounds, like speech, instructions, or directional cues, while cutting through background noise. 

\textit{Mid Range.} Roles like \textbf{interpret} (9.8\%) and \textbf{guide} (8.8\%) showed relatively balanced contributions across both datasets. By contrast, \textbf{anticipate} (8.4\%), and \textbf{simulate} (7.0\%) appeared more prominently in the autoethnography than in the survey. These roles reflect participants’ interest in AAR as a perceptual assistant, capable of interpreting complex soundscapes, providing early warnings, and offering situational prompts to support movement and attention.

\textit{Lowest}. Roles such as \textbf{transform}, \textbf{remember}, and \textbf{reflect} appeared across both datasets but remained consistently low in frequency. Rather than indicating a divergence between methods, this pattern reflects a shared tendency: these roles occupy the speculative and affective end of the design space and were less commonly raised in both the CAE and survey. The autoethnography surfaced them slightly more often, but this reflects the reflective depth afforded by lived engagement rather than a substantive difference in priority.

\textit{Current vs. Future-Facing Roles.} The three roles that correspond to existing capabilities together make up 47.5\% of all coded instances. Participants in the survey drew more heavily on current auditory practices (53.2\%) compared to the CAE (36.8\%). Broader users anchored their expectations in familiar capabilities, while the autoethnography surfaced a comparatively wider spread of future-facing possibilities.

\begin{figure}[ht]
  \centering
\includegraphics[width=1\linewidth]{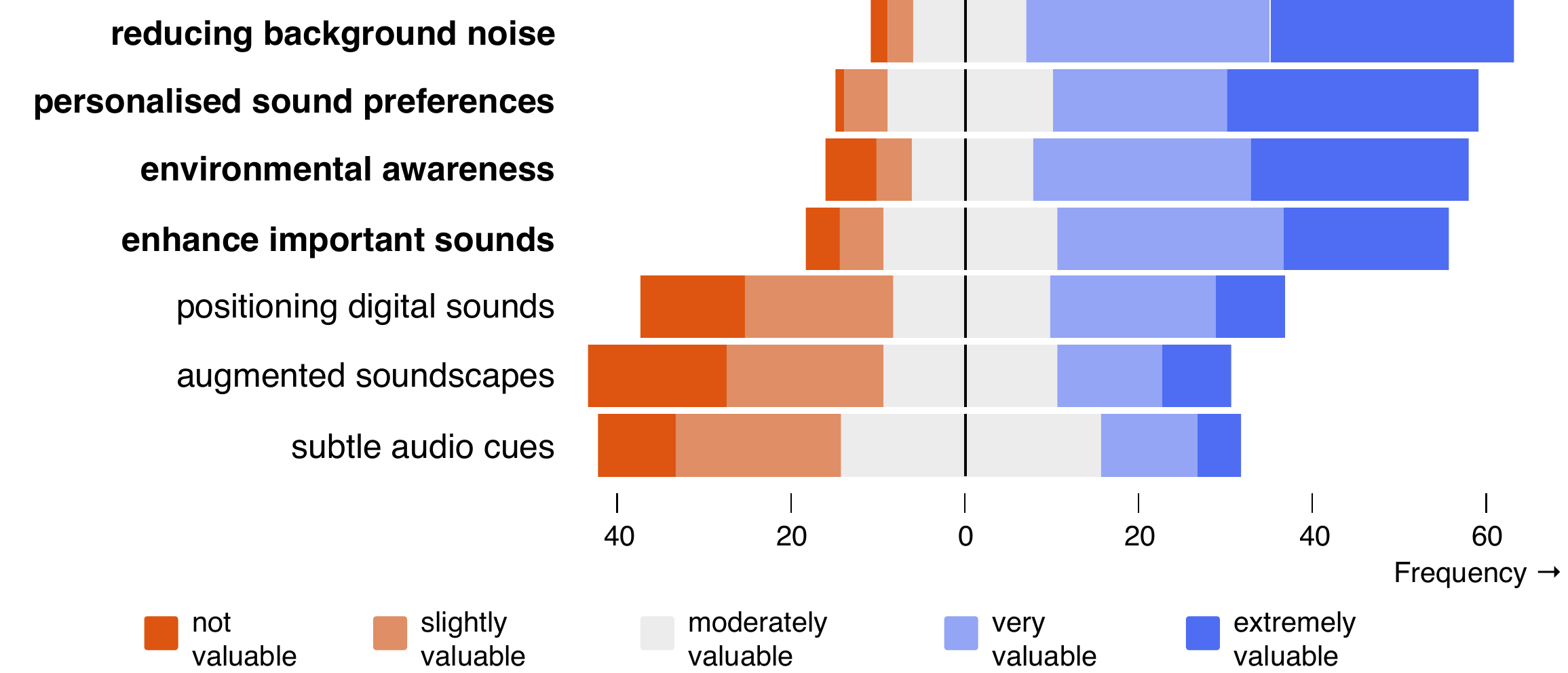}
  \caption{Survey participant ratings of AAR features. Responses to Q7: \textit{`If AAR could transform sound in real time, which of the following features would be most valuable to you?'} }
  \label{fig:aar_features}
  \Description{The figure presents survey participant ratings of seven potential AAR features in response to the question: “If AAR could transform sound in real time, which of the following features would be most valuable to you?” Each feature is displayed as a horizontal diverging bar chart, with responses ranging from “not valuable” (dark orange) to “extremely valuable” (dark blue). The most highly rated features, with strong lean towards “very” and “extremely valuable,” are reducing background noise, personalised sound preferences, environmental awareness, and enhancing important sounds. Features such as positioning digital sounds, augmented soundscapes, and subtle audio cues received more mixed responses, with higher proportions of “slightly” or “moderately valuable” ratings. Frequency is plotted on the horizontal axis, with positive values indicating higher-value ratings and negative values showing lower-value ratings.}
\end{figure}

These open-ended responses were complemented by a separate question (Q7) asking survey participants to rate the value of specific real-time sound transformation features. As shown in \autoref{fig:aar_features}, the highest-rated features included \textit{reducing background noise}, \textit{personalised sound preferences}, \textit{environmental awareness} and \textit{enhancing important sounds}, with the majority of participants selecting these as either \textit{very} or \textit{extremely valuable}. These closely align with the most frequently coded roles in the open responses. 

\subsubsection{Control Preferences and Everyday Concerns}
\label{survey-concerns}

Responses to Q10 revealed strong preferences for maintaining some level of control over AAR systems. Most participants agreed with the ability to manually adjust settings or to set preferences once and allow the system to adjust within those limits. While a majority also supported more adaptive modes, such as adjusting based on implicit actions (e.g., gaze, head movement) or fully adapting in the background, these options showed a broader spread of responses (see \autoref{fig:aar_control}). Open-ended reflections (Q11) revealed five recurring motivations:

\begin{figure}[ht]
  \centering
\includegraphics[width=1\linewidth]{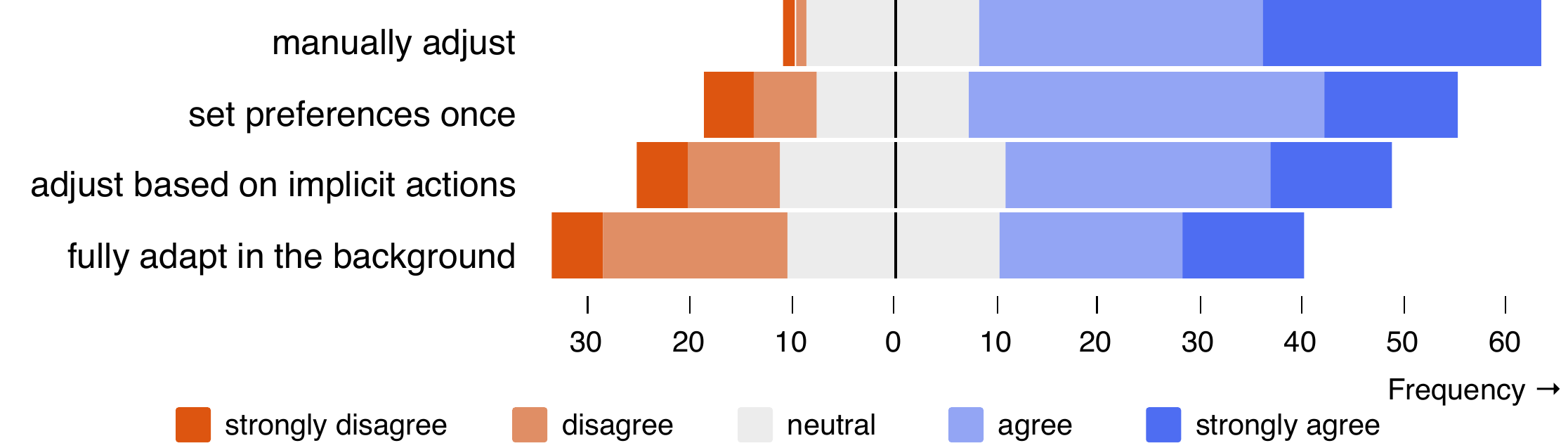}
  \caption{Survey participant preferences for controlling AAR. Responses to Q10: \textit{`How much control would you want over AAR?'} }
  \label{fig:aar_control}
  \Description{The figure shows survey participant preferences for controlling AAR in response to the question: “How much control would you want over AAR?” Results are displayed as a horizontal diverging bar chart for four control modes. Responses range from “strongly disagree” (dark orange) to “strongly agree” (dark blue). The most preferred approach is manually adjust, with a strong majority agreeing or strongly agreeing. Set preferences once also receives high agreement, though slightly less than manual adjustment. Adjust based on implicit actions shows a more balanced split, with many neutral responses. Fully adapt in the background receives the most disagreement overall, though it still has notable agreement among some participants. Frequency is shown on the horizontal axis, with positive values representing agreement and negative values representing disagreement.}
\end{figure}

Many participants expressed a clear and deliberate preference for manual control over AAR. For this group, control was not just a convenience, it was essential for achieving \textbf{a sense of reliability and personal fit}. Participants frequently referred to past frustrations with automation and highlighted how current technologies fail to deliver truly personalised experiences, noting that automated systems \textit{`never get it quite right'} and often miss subtle, day-to-day variations in needs. As one participant put it, \textit{`analogous technologies are not great,'} citing examples like screen brightness adjustments that frequently fall short. Manual control was also seen as the key to achieving an optimal experience: \textit{`I am just after quality and maximum experience.'} Others valued the ability to \textit{`test and play around with the settings'} to discover what works best and adapt the system to their needs.

Another distinct motivational layer behind participants’ desire for manual control was the need to \textbf{preserve personal agency and resist unwanted influence}. This included concerns that automation might make decisions on their behalf. As one participant put it, \textit{`I feel that if I am not controlling an item, then someone else is,'} while another simply expressed \textit{`concerns around the AR making choices for me.'}

A number of participants expressed \textbf{caution toward implicit control}, questioning whether behaviour-based adjustments would be accurate or genuinely helpful. As one put it, \textit{`The adjustment based on my actions seems a bit concerning... it might be useful, or just annoying.'} Others were unsure whether such systems would act in ways aligned with their real-time needs or preferences. Privacy was also a recurring concern in this cluster. Statements like \textit{`I wouldn’t want my devices to monitor my behaviour all the time'} reflect discomfort with continuous sensing, even when framed as a convenience.

Several participants expressed a desire for \textbf{user-configurable intelligence}. Rather than rejecting adaptive features, they emphasised the need for systems that can learn from user input and support override when needed. One participant suggested, \textit{`I would like it to automatically adapt but be responsive to manual corrections, remember them, and learn from them,'} while another noted they wanted to \textit{`set preferences once... but also manually control it if things need tweaking.'} Finally, a subset of participants emphasised the importance of \textbf{minimising cognitive and interactional effort} in controlling AAR. For these users, the ideal system would be \textit{`smarter, more adaptive to the environment and mood,'} requiring little or no ongoing input. Several noted that they {`don’t want to bother setting it up or modifying settings all the time,'} preferring systems that can learn preferences once and \textit{`just work.'} 

\begin{figure}[ht]
  \centering
\includegraphics[width=1\linewidth]{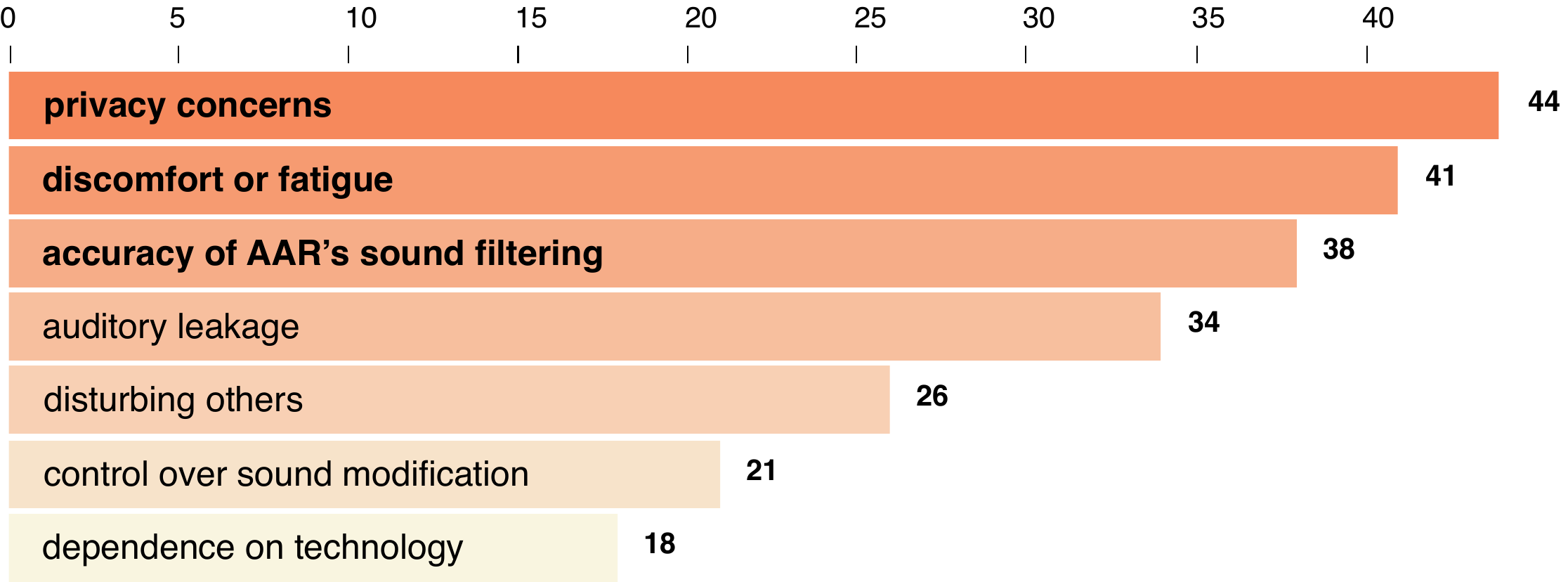}
  \caption{Top three concerns selected by each survey participant about using AAR in everyday life. Responses to Q12: \textit{`What concerns would you have about using AAR in everyday life?'}}  
  \label{fig:aar_concerns}
  \Description{The figure shows the top three concerns selected by each survey participant about using AAR in everyday life, in response to the question: “What concerns would you have about using AAR in everyday life?” The horizontal bar chart lists seven concerns, with bar lengths indicating the number of participants selecting each. Privacy concerns (44) is the most frequently selected, followed by discomfort or fatigue (41) and accuracy of AAR’s sound filtering (38). Other concerns include auditory leakage (34), disturbing others (26), control over sound modification (21), and dependence on technology (18). Bars are shaded from darker orange for the most frequently cited concerns to lighter tones for less frequent ones.}
\end{figure}

Meanwhile, responses to Q12 highlighted common concerns such as privacy, fatigue, sound filtering accuracy, and the risk of disturbing others, many of which intersect directly with participants' control preferences. Fewer participants raised issues around dependence on technology or lack of control over sound modification (see \autoref{fig:aar_concerns}). 

\section{Discussion}

This section examines how participants imagined integrating AAR into everyday life (RQ1) and the concerns that arose in the process (RQ2) and their implications for future design.

\subsection{Envisioned Integrations of AAR (RQ1)} 

Prior research on AAR has primarily focused on task- or domain-specific applications. For example, a systematic review by~\cite{yang2022audio} categorised AAR applications into seven domains: navigation and location-awareness assistance, augmented environmental perception, presentation and display, entertainment and recreation, telepresence, education and training, and healthcare. While these domains reflect the diversity of current research efforts~\cite{rasheed2024critical, ruminski2015experimental, chen2022lisee, li2024beyond}, they offer limited integration insights into everyday life. In contrast, our study examined how people imagine AAR shaping ordinary routines, surfacing ten roles that offer a more grounded and user-centred vision of audio augmentation.

\subsubsection{Task- and utility-oriented roles}

Several roles surfaced in our study, such as \textbf{enhance}, \textbf{reduce}, and \textbf{guide} closely align with task- or utility-oriented applications previously described in AAR research. These include domains such as augmented environmental perception and navigation support~\cite{yang2022audio}, as well as more recent work on everyday audio augmentation that foregrounds user control, selective filtering, and comfort in shared or noisy environments~\cite{mathis2024everyday, bustoni2024exploring, jacobsen2024towards}. The recurrence of these roles in both the autoethnography and the survey suggests that functional and context-sensitive auditory support remains a salient part of how people imagine AAR in daily life, even when prompted to speculate more broadly.

\subsubsection{Emotional and social roles}

While some roles aligned with task-oriented uses, others pointed to how AAR might support more emotional or socially nuanced engagement. For instance, roles such as \textbf{personalise}, \textbf{anticipate}, \textbf{remember}, and \textbf{reflect} suggest that AAR could enrich subjective and affective experience rather than merely solve practical problems. These roles echo the MemoryMR and CommunicateMR categories proposed by \citet{mathis2024everyday}, yet while prior work has primarily addressed these themes in the context of multimodal AR, our findings suggest that audio alone can carry subtle emotional and social weight. Our audio-centred approach also revealed subtleties that may be less visible in multimodal frameworks. Participants described becoming hyper-aware of how their own sounds (e.g., chewing, cleaning, typing) might be perceived by others; others imagined systems that could respond to emotional states. By modulating experience peripherally, AAR is particularly suited to supporting emotional attunement in subtle, continuous ways. These comparisons suggest that while AAR shares common ground with broader AR visions, it also introduces distinctive sensory and interpretive affordances.

\subsubsection{Perceptual collaborator roles}

Beyond emotional attunement, a smaller but compelling set of \textbf{interpret} envisioned AAR as more than a sensory aid, imagining it as a perceptual collaborator that supports how people interpret experiences and relate socially through sound. A sensory enhancer might help a user hear a fire alarm more clearly; a perceptual collaborator could help them understand what kind of alarm it is, why it matters, and whether they need to respond. 
This desire for sound awareness is echoed in emerging technological visions, such as Meta’s \textit{`auditory machine perception'}~\cite{meta2020audioresearch}, which frames the opportunity not only in helping us hear better, but in helping us understand better. Such a view aligns with perspectives from cognitive science and embodied interaction theories. As McLuhan argued, \textit{`all media are extensions of some human faculty, psychic or physical'}~\cite{mcluhan1994understanding}, headphones extend the ear, but AAR reshapes how we listen and interact. Similarly, the \textit{`extended mind'} thesis~\cite{clark1998extended} suggests that cognition is not confined to the brain but distributed across tools and environments. From this view, AAR becomes a medium for extending perception, not just altering the soundscape, but changing how users engage with and make sense of their world.

Of note, these varied roles were not uniformly distributed across our two methods (see \autoref{tab:aar_percentages}). Survey responses tended to emphasise practical and immediate needs, such as comfort, personalisation, and perceptual clarity, reflecting common challenges and user expectations. In contrast, the CAE surfaced more reflective and speculative roles, exploring how AAR might shape meaning, memory, or social presence. These patterns reveal a valuable triangulation: while broader participants focused on what AAR should do, the researcher reflections opened up what AAR could mean.

\subsubsection{Rhythmic and embodied collaborator}

Beyond these roles, we can situate AAR within broader HCI theory to foreground its entanglement with everyday temporality and embodied practice. Drawing on Lefebvre’s rhythmanalysis~\cite{lefebvre2013rhythmanalysis} and Dourish’s embodied interaction~\cite{dourish2001action}, we propose that AAR can be understood as a rhythmic collaborator that mediates how people live through and make sense of time, sound, and social space. 

Lefebvre distinguishes between rhythms at multiple scales: micro-rhythms (the body’s movements and sensations), meso-rhythms (situated routines and activities), and macro-rhythms (wider cycles such as workdays, commutes, and seasonal flows). Positioning AAR through rhythmanalysis clarifies why the ten roles do not stand as isolated functions but as interventions distributed across the temporal structure of everyday life. For instance, \textbf{reflect} engage micro-rhythms by attuning users to the sound of their own breathing, chewing, or footsteps. \textbf{Anticipate}, \textbf{remember}, and \textbf{simulate} align with macro-rhythms, linking auditory experiences to long-term temporal cycles such as waiting for daily deliveries, marking seasonal soundscapes, or recalling cultural memories. Others intervene at the level of meso-rhythms, supporting focus in work, commuting, or domestic routines (see \autoref{tab:aar_rhythms}).

From an embodied interaction perspective, these rhythms are not abstract but lived through the body’s orientation, attention, and movement. For example, selectively enhancing a voice in a crowded space changes how a body turns toward speakers; transforming intrusive environmental noise into calming audio reconfigures affective states and postures; guiding footsteps with spatial cues alters navigational patterns. Such examples underscore AAR’s potential to redistribute sensory engagement in ways that are felt physically as much as cognitively.

This framing also has practical implications. Seeing AAR as acting across different rhythms highlights that design choices need to match the timing and structure of everyday life. For example, at the micro-scale, this may involve providing a gentle pacing cue during a walk that helps maintain a steady rhythm without masking surrounding sounds. At the meso-scale, designs should align with ongoing routines; for example, a commuting aid that enhances traffic cues should do so without repeatedly overriding the music a user relies on during the trip. At the macro-scale, systems should remain sensitive to longer-term patterns, such as daily delivery notifications or seasonal soundscapes.

\begin{table*}[ht]
\centering
\small
\caption{Mapping AAR roles onto Lefebvre’s rhythmic scales.}
\label{tab:aar_rhythms}
\begin{tabular}{@{}p{2.7cm} L{6cm} L{4cm}@{}}
\toprule
\textbf{Rhythmic Scale} & \textbf{Examples of Everyday Life} & \textbf{AAR Roles} \\
\midrule

\textbf{Micro-rhythms} 
& \textit{Bodily cycles:} \newline breathing, footsteps, chewing, typing 
& reflect \\[3pt]

\textbf{Meso-rhythms} 
& \textit{Situated routines:} \newline commuting, cooking, working, socialising 
& reduce, enhance, guide, interpret, personalise, transform \\[6pt]

\textbf{Macro-rhythms} 
& \textit{Broader cycles:} \newline workdays, seasonal / environmental soundscapes 
& anticipate, remember, simulate \\

\bottomrule
\end{tabular}
\end{table*}

\subsection{Concerns Around AAR Integration (RQ2)}

Our study surfaced overlapping but distinct sets of concerns about AAR, shaped by the differing nature of the survey and CAE. Survey participants raised both \textit{system-level concerns}, centred on privacy, control, and automation, and more experiential issues related to fatigue (see Section~\ref{survey-concerns}). Participants expressed unease about their AAR devices potentially listening at all times, capturing conversations or personal moments without clear indication or consent. This concern was also reflected in the dislike for implicit control, which was perceived as requiring the system to constantly monitor user behaviour. These anxieties around data collection, surveillance, and unintended recording point to a need for transparent affordances and user control, ensuring that people can clearly understand and manage when, what, and how sound is captured in everyday use. We also note that in our study, concerns centred primarily on the user’s own privacy rather than bystander privacy—a contrast to visual AR, where camera-enabled smart glasses raise stronger worries about recording others~\cite{o2023privacy, tran2025everyday}. One possible reason is that audio capture may be perceived differently from video capture, whether in terms of visibility, intrusiveness, or perceived consequence.

CAE participants, in contrast, reflected on how audio augmentation might create \textit{subtle tensions} in their everyday lives (see Section~\ref{tensions}). Their reflections collectively reveal that everyday listening supports several things at once, such as staying aware of one’s surroundings, picking up social cues, monitoring safety, and maintaining an emotional sense of place. Because these functions operate simultaneously, modifying any part of the soundscape can produce knock-on effects across them. This helps explain why seemingly small augmentations, such as suppressing distractions or spatialising voices, raised concerns: filtering noise might dull social awareness, personalised spatial effects could shift the emotional feel of an interaction, and reducing ambient sound risks obscuring cues needed for safety. These concerns differ from system-focused worries reported in prior work, such as manipulation through overlays~\cite{mathis2024everyday}, and align with \citet{bustoni2024exploring}, who found that users favour lighter augmentation when ambient sound carries informational value.

Our findings reflect two complementary levels of concerns: survey responses pointed to issues of \textit{system trust, data use, and control}, while CAE reflections revealed how AAR might interfere with \textit{attention, comfort, or interpersonal rhythm}. This layered perspective underscores the importance of designing AAR not only to protect user rights, but also to remain sensitive to the lived and often ambiguous nature of auditory engagement.

Such findings refine broader critiques of augmentation. \citet{mathis2024everyday}, for example, describe the `dark side of augmenting humans' in terms of system-level risks such as manipulation and over-reliance. \citet{de2022deceiving} similarly highlight how manipulative audio design, such as spatial cues that mislead movement, may undermine user agency. Our findings extend this critique by surfacing more situated and sensory-specific risks. In this sense, the `dark side' of AAR may be about misalignment, when audio augmentation interferes with how people naturally manage attention, comfort, and connection in everyday life. These concerns may also help explain the strong preference of the survey participants to retain manual control over AAR systems. As audio augmentation operates in intimate and often ambiguous sensory spaces, participants expressed a desire for AAR systems that offer precision and adjustability, not because they reject automation, but because the sensory stakes are so intricate. 


\subsection{Designing for Emerging and Speculative AAR Roles}

While many of the AAR roles described by participants reflect extensions of current smart audio capabilities, such as spatial audio~\cite{blauert1997spatial, sundareswaran2003aar, kailas2021spatial}, others point to more speculative directions that raise important design and technical challenges. For instance, the role of \textbf{reflect} suggests systems that could help users understand how their own sound production is perceived by others. Realising this would require not just environmental sensing, but the ability to model social expectations and attentiveness. While some work in wearable computing explores proxemic audio cues or gaze-aware interfaces~\cite{Laporte2021detecting, Bulling2008eyes}, socially attuned audio feedback in open environments remains an open challenge.

Similarly, roles like \textbf{interpret} and \textbf{remember} involve real-time interpretation of ambient sound. Advances in sound recognition, such as Google’s Sound Notifications~\cite{Savla2020household} or Apple’s Sound Recognition~\cite{Apple2025recognise} for accessibility, demonstrate that detecting alarms, appliances, or pet sounds is possible in controlled conditions. However, dynamically annotating or contextualising complex or subtle audio (e.g., mechanical issues, background languages) is still at the edge of what’s possible in mobile form factors. Participants’ interest in emotionally adaptive soundscapes also raises questions about affective sensing. While recent work has explored biofeedback-driven auditory feedback to support body awareness and meditative movement~\cite{pang2024wearable, tan2025runme}, aligning audio augmentation with a user’s internal state, without feeling manipulative or `too knowing', requires more work in multimodal inference and user calibration.

Even roles that seem technically feasible, such as \textbf{enhance}, present subtle challenges when applied in real-world settings. Participants often imagined selectively amplifying certain sounds, such as hearing their name called at a café and distinguishing voices in group settings. However, current systems still struggle to determine which sound the user actually wants to enhance in the moment. Approaches like directional microphones or beamforming can improve signal-to-noise ratios in known contexts~\cite{meta2020audioresearch}, and recent research in attention-guided hearing aids has explored using gaze direction or brain activity (e.g., EEG~\cite{van2016eeg}) to infer auditory focus. Yet these methods are either limited in mobility, require intrusive sensing, or depend on narrow interaction windows. In less structured environments, accurately identifying the `target' sound remains technically challenging. 

These findings suggest that while AAR invites playful and meaningful design opportunities, many envisioned roles will require new technical infrastructures and more responsive models of context, emotion, and sociality. 

\subsection{AAR Devices and Form Factors}

AAR can be supported across a range of device form factors. Spatial audio features in earbuds such as Apple AirPods Pro, Samsung Galaxy Buds Pro, and JBL Quantum ONE offer spatialisation of virtual sounds~\cite{yang2022audio}. Headsets like the Apple Vision Pro\footnote{\url{https://www.apple.com/au/apple-vision-pro/}}
 integrate spatial sound within broader immersive environments. Drawing on workshop findings from \citet{dam2022extracting, dam2024taxonomy}, air- and bone-conduction earphones are currently the most preferred wearables for AAR, while smart glasses remain comparatively less popular.

AAR glasses inherit many of the social challenges of visual smart glasses, particularly ambiguity around attention and engagement. Unlike headphones, they offer no visible cues that the wearer is listening to augmented audio. The result is a hybrid form factor that is technically unobtrusive but socially opaque. Because AAR use occurs in shared social environments, this opacity affects both sides of an interaction: others cannot easily tell whether the wearer is attentive or partially occupied, and the wearer cannot rely on established signalling conventions (such as visible headphones) to indicate availability or disengagement. Our findings suggest that the design of AAR form factors must therefore consider not only acoustic performance but also how they make presence and attentional shifts socially legible.

A second form-factor consideration concerns audio leakage. Unlike sealed headphones, many AAR devices rely on open-ear or bone-conduction designs, which can unintentionally project snippets of augmented sound into shared spaces. Participants noted that this undermines the assumed privacy of AAR and may breach social norms around quietness or confidentiality. Whereas the opacity of AAR glasses makes attentional state difficult to read, audio leakage introduces the opposite problem: unintended outward signalling. These effects highlight that AAR form factors must negotiate not only what the wearer hears, but also what the environment hears, making acoustic containment an important design concern.

\subsection{Methodological Reflections}

This study offers insights into everyday visions of AAR through a dual-method approach. A key strength of our approach lies in its framing of sound not only as a source of friction or disruption but also as a site of opportunity. Rather than focusing solely on problems to be solved, we deliberately encouraged participants to reflect on positive, meaningful, or resonant auditory experiences. This orientation helped surface design opportunities that build on what already works: moments of calm, emotional connection, social rhythm, or personal significance. In doing so, our analysis attends not just to what needs fixing, but to what might be transformed.

At the same time, this study has several limitations. First, the autoethnographic reflections were authored by a small group of researchers with shared disciplinary backgrounds. Although these reflections allowed for depth and situated insight, they may not represent the full range of lived experiences, particularly across different cultural or ability contexts. A further methodological consideration concerns our use of CAE. While we draw on CAE’s emphasis on first-person reflection and dialogic interpretation, our approach represents a pragmatic adaptation~\cite{kaltenhauser2024playing}. Rather than engaging in full cultural critique or autobiographical narration, we used CAE to structure collaborative, reflexive engagement with our everyday auditory experiences. In this sense, our method sits between CAE and reflexive diary studies. Reflexive diary studies rely on private, individual reflection, whereas our approach involved shared, iterative dialogue among researcher-participants. This collaborative reflection extends beyond diary methods while remaining short of full autoethnographic cultural critique.

Second, the survey sample size (N=74) was appropriate for an exploratory mixed-methods investigation in an emerging research area, though relatively modest compared to larger-scale quantitative studies~\cite{bustoni2024exploring, abraham2024record}. Future work could build on this foundation through broader surveys to strengthen generalisability. Additionally, although the sample was demographically diverse, participant engagement varied in depth. Some responses were brief or lacked the experiential grounding present in the autoethnography. That is, survey participants were typically imagining potential uses for AAR rather than reflecting on direct experiences or lived engagement. These differences reflect the kinds of reflection each method supported. Future studies could bridge these layers through methods that scaffold more sustained or situated reflection, such as cultural probes~\cite{yu2024probe} or diary studies~\cite{li2024diaryhelper}.

Third, both components of the study relied on imagined or retrospective accounts rather than real-time interactions with working AAR systems. As such, the findings reflect user desires and interpretive framings, rather than observed behaviours or embodied experiences. Future work should explore these directions through longitudinal or prototype-based studies, which would be better suited to generating the empirical evidence required to develop design guidelines.

Finally, our study focused exclusively on audio-based use cases and did not examine scenarios in which audio complements visual or multimodal AR experiences. This was a deliberate choice aimed at foregrounding the unique affordances, challenges, and roles of AAR in everyday life, independent of visual augmentation. Prior work highlights the value of isolating audio as an independent modality to better understand its specific contributions to immersive experience~\cite{hedges2023measuring}. However, as many AR systems increasingly combine visual and auditory modalities, future research should investigate how audio interacts with or enhances visual content, and how this integration might shift user expectations, roles, and concerns. For example, studies on audio-visual latency~\cite{hopkins2022latency} and comparisons of spatialised cues~\cite{barde2019less, zhang2023see} suggest that the relationship between modalities is complex and context-dependent. While \cite{zhang2023see, schwandt2024audiovisual} found that combining visual and audio cues can enhance AR task performance, \cite{barde2019less} showed that visual and auditory cues can perform better on their own. Other work shows that audiovisual distractors can impair auditory attention~\cite{moraes2024distractor}.

\section{Conclusion}

As audio technologies become more embedded in everyday life, there is a need to understand how people imagine their integration not only functionally but experientially. This study surfaces how people envisage AAR supporting everyday listening, identifying ten roles through grounded accounts of lived sound experience. The roles align with Schraffenberger’s \textit{Arguably AR} thesis~\cite{schraffenberger2018arguably} by framing augmentation as an experience-driven phenomenon grounded in the relationship between the virtual and the real, and not limited to visual overlays. At the same time, participants raised important concerns. They questioned the appropriate level of control, the privacy implications of constant audio capture, and whether adaptive systems can ever fully align with complex, shifting human needs.

Rather than starting from technical capability, the study begins from what people report needing and noticing in their environments. Our contribution differs from industry visions in two ways. First, the roles derive from fine-grained accounts of everyday listening rather than from speculative product narratives. Second, our analysis foregrounds the experiential considerations that shape how people imagine augmentation fitting into daily life, aspects that are typically simplified or absent.

\begin{acks}
The authors thank the anonymous reviewers for their valuable comments and suggestions, which have been incorporated into the final version.
\end{acks}

\bibliographystyle{ACM-Reference-Format}
\bibliography{references}

\appendix

\section{Declaration of AI Use}
\label{sec:declaration}

ChatGPT was used to generate the illustrations presented in \autoref{fig:aar_roles}. The following prompts were provided for each AAR role:

\begin{itemize}
    \item \textbf{Enhance}: `A simple hand-drawn illustration of a smiling person cupping their ear, with clear sound waves (blue) reaching them. Bold black outline, minimal detail, playful style.'
    \item \textbf{Reduce}: `A simple hand-drawn illustration of a person happily relaxing while a loudspeaker nearby is crossed out (blue slash). Bold black outline, playful style.'
    \item \textbf{Guide}: `A simple hand-drawn illustration of a person standing with an arrow (blue) and sound waves guiding them forward. Bold black outline, playful style.'
    \item \textbf{Personalise}: `A simple hand-drawn illustration of a person adjusting sliders (blue) on a floating audio panel. Bold black outline, playful style.'
    \item \textbf{Anticipate}: `A person in a room looks up at a small bell with soft blue waves coming out gradually, as if the volume is ramping up instead of suddenly blasting.'
    \item \textbf{Interpret}: `A simple hand-drawn illustration of a person listening with curiosity as a speech bubble above them turns into text (blue). Bold black outline, playful style.'
    \item \textbf{Remember}: `A simple hand-drawn illustration of a person smiling as they hear a sound that triggers a thought bubble (blue) showing a memory scene. Bold black outline, playful style.'
    \item \textbf{Simulate}: `A simple hand-drawn illustration of a person walking with musical notes (blue) floating around them, looking cheerful. Bold black outline, playful style.'
    \item \textbf{Reflect}: `A simple hand-drawn illustration of a person making a sound (like chewing or typing), with blue sound lines coming out, and the person looking self-aware or embarrassed about it. Bold black outline, playful style.'
    \item \textbf{Transform}: `A simple hand-drawn illustration of a person listening to jagged sound waves that gradually change into musical notes (blue). Bold black outline, playful style.'
\end{itemize}

\section{Study Data}
\label{sec:studydata}

The CAE data, survey data, and survey questions have been published on the Open Science Framework (OSF): \url{https://osf.io/8hszk} As the CAE data contains highly personal reflections, the researchers’ real names have been replaced with pseudonyms.

\section{Online Survey Distribution}
\label{sec:online}

\begin{table*}[ht]
\caption{Online communities where the survey was distributed.}
\centering
\small
\begin{tabular}{p{1.5cm}L{2.5cm}p{6cm}}
\toprule
\textbf{Platform} & \textbf{Community} & \textbf{URL} \\
\midrule
Reddit & r/headphones & \url{https://www.reddit.com/r/headphones/} \\
       & r/augmentedreality & \url{https://www.reddit.com/r/augmentedreality/} \\
       & r/SmartGlasses & \url{https://www.reddit.com/r/SmartGlasses/} \\
       & r/RayBanStories & \url{https://www.reddit.com/r/RayBanStories/} \\
       & r/RaybanMeta & \url{https://www.reddit.com/r/RaybanMeta/} \\
\midrule
Facebook & Ray-Ban Meta Smart Glasses Community & \url{https://www.facebook.com/groups/metasmartglasses} \\
         & Blind Users of Meta Glasses & \url{https://www.facebook.com/groups/5930139520442990} \\
\bottomrule
\end{tabular}
\label{tab:survey_distribution}
\end{table*}

\label{sec:data}

\end{document}